\documentclass[conference]{IEEEtran}

\usepackage{amsmath}
\usepackage{cite}
\usepackage[margin=26mm]{geometry}
\usepackage[final]{graphicx}
\usepackage{bm}
\usepackage{amssymb}
\usepackage{color}
\usepackage{tikz}
\usetikzlibrary{shapes.geometric}

\usepackage{epsfig}
\usepackage{nicefrac}
\usepackage{algorithm}
\usepackage[noend]{algpseudocode}

\title{Network Optimization - \\Using Relays as Neurons}

\author{\IEEEauthorblockN{Itsik Bergel}
\IEEEauthorblockA{Faculty of Engineering\\
Bar Ilan University\\
Ramat-Gan, Israel\\
Email: itsik.bergel@biu.ac.il}
}

\date{April 2023}

\begin{document}

\maketitle

\begin{abstract}
We consider the optimization of a network with amplify-and-forward relays.  Observing that each relay has a power limit, and hence a non-linear transfer function, we focus on the similarity between relay networks and neural networks.  This similarity allows us to treat relays as neurons, and use deep learning tools to achieve better optimization of the network. Deep learning optimization allows relays to work in their non-linear regime (and hence increase their transmission power) while still avoiding harmful distortion. Moreover, like neural networks, which can implement almost any functionality,  we can take advantage of the non-linearities and implement parts of the received functionalities over the relay network. By treating each relay element as a node in a deep neural network, our optimization results in huge gains over traditional relay optimization, and also allows the use of simpler receivers.
\end{abstract}
\section{Introduction}

Relays have been essential for improving communication performance since the early days of wireless communication (e.g., \cite{harmon1912girdling,cover1979capacity,laneman2004cooperative}). 
Deploying more relays is a key to meeting the exponentially growing demand for wireless communication. 
 However, the optimal operation of large relay networks is still infeasible. 

 This research will focus on amplify-and-forward relays (e.g., \cite{nabar2004fading,zhao2006improving}), which are simple to design and optimize. In particular, these relays became more accessible by recent  progress in two technologies: Energy harvesting (e.g., \cite{nasir2013relaying,lu2014wireless}) and Full duplex communication (e.g., \cite{riihonen2011mitigation,sabharwal2014band}). 

Optimization of amplify-and-forward relay networks is far from trivial, as their performance  usually presents a non convex behavior. Nevertheless, modern optimization methods know  to converge to at least a local maximum for many relay networks \cite{good_phan2012beamforming,sanguinetti2012tutorial}.

 However, all traditional optimization of relay networks consider the relays as linear amplifiers with a power constraint. Thus, the use of such methods must limit the relay operation to the regime where it can be approximated as linear. As a result, these methods are forced to set a power constraint that is lower than the actual power achievable by the relay.

In this work, we take a completely different approach. We observe that each relay naturally has a power limit and hence has a non-linear transfer function. Thus, we focus on the similarity between a relay network and a neural network.

Neural networks have gained much popularity in recent years due to their ability to solve tough computational challenges, and in particular were good modeling of the problem is not available. The use of neural networks was suggested in many communication applications (e.g., \cite{chen2019artificial,nikbakht2020unsupervised ,sholev2020neural  
,saxena2021reinforcement}) and even in relay applications (e.g., \cite{zhang2020neural,xu2021intelligent,guo2021energy
}).
In particular   \cite{zhang2020neural} used neural networks for relay selection, \cite{guo2021energy} combined it also with power allocation, and \cite{xu2021intelligent
} used a neural network for the prediction of outage probabilities.

In this work, we use neural network technology, but in a completely different way.  The resemblance between a non linear relay and a neuron  
allows us to treat relays as neurons, and use deep learning tools to achieve better optimization of the relay network. Moreover,  neural networks can implement almost any functionality. Thus, through proper training, we can take advantage of the non-linearities and implement parts of the receiver functionalities over the relay network thus reducing the receiver complexity.

Preliminary results show a huge $25$dB improvement over the state of the art, in a cellular network with 100 relays and 2 users. These results also demonstrate the ability of the relay network to non-linearly separate the signals for the two receivers.

The main advantages of the proposed scheme are:
\begin{itemize}
\item Better communication over relay networks.
\item New computational capabilities ``over the air''.
\item Support for distributed optimization.
\end{itemize}

The main differences of the proposed scheme from other implementation of neural networks:
\begin{itemize}
\item No added neurons - Relays are treated as neurons but are an actual part of the network. 
\item Limited control on the network topology - Most of the network topology is determined by the channel gains, which result from physical phenomena. The network optimization can only control the gain (and possibly bias) at each relay.

\item Noisy ``neurons'' - The input of each relay suffers from additive noise. Thus, we need to cope with many noise sources within the network.
\end{itemize}

In the following, we first present the system model in Section II. Section III briefly 
discusses the traditional optimization approach, while Section IV presents our novel deep learning approach.  Section V, presents numerical studies that demonstrate the advantages of our approach and Section VI gives our concluding remarks.

\begin{figure}[t]
 \centering
\epsfig{figure=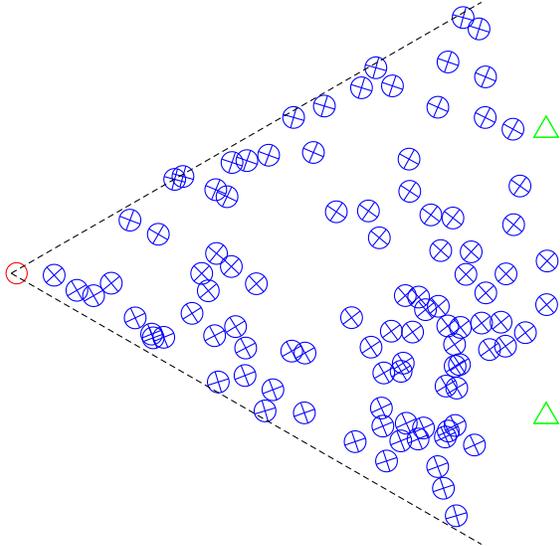,trim=0cm 0cm 0cm 0cm, clip=true,width=0.45\textwidth}
\caption{\label{fig:system_model} 
A relay network example with a transmitter (red circle) $M=2$ receivers (green triangles) and $N=100$ relays (blue circles with x marks). 
}
\end{figure}

\section{System model}
We consider a network with a single transmitter (e.g., base station), $N$ amplify-and-forward relays and $M$ receivers, as demonstrated in Fig \ref{fig:system_model}. All transmissions in the network are performed over the same frequency, and all relays work in full-duplex mode.. The transmitter simultaneously transmits independent data to each of the receivers. 

For  clarity, this basic study of a deep relay network takes two simplifying assumptions:  We assume that all signals are real (i.e., transmitted signals, channel gains and relay gains are all real)  and we assume perfect full duplex and directional antennas. These assumptions significantly simplify the mathematical presentation while still keeping the essence of the relay network. Both assumptions will be relaxed in future studies.    

Dropping the time index, the bits intended for receiver $m$ are denoted by $\mathbf{u}_m=[u_{m,1},\ldots,u_{m,K}]$ where $K$ denotes the number of bits simultaneously transmitted to each receiver.
We consider a single antenna transmitter (and hence, signal separation cannot be done by spatial multiplexing). Thus, all transmitted data is jointly modulated to a single symbol. This is done by stacking the bits intended for all users into a single vector, $\mathbf{u}=[\mathbf{u}_1,\ldots,\mathbf{u}_M]$, using gray code and then pulse amplitude modulation (PAM). Without loss of generality, the maximal absolute value at the transmitter output is set to $1$. For example, for $K=1$ and $M=2$, the $4$ PAM points to deliver one bit per channel use per user are shown in Table I.

\begin{table}[h]
\label{T:Table 1}
\begin{center}
\begin{tabular}{ |c|c|c|c|c| } 
 \hline
 Bit for user 1, $u_{1,1}$ & 0 & 0 & 1 &1 \\ 
 Bit for user 2, $u_{2,1}$ & 0 & 1 &1 &0\\ 
 \hline 
 Transmitted value & -1 & -1/3&1/3&1 \\ 
 \hline
\end{tabular}
\end{center}
\caption{Constellation points for transmitting 1 bit per users for 2 users.}
\end{table}
\vspace{-5mm}
The amplify-and-forward relay has limited power. Traditional analysis considers the relays as linear amplifiers with a power constraint. To improve performance, we consider the actual transfer  function of the relays.

\begin{figure}[t]
\epsfig{figure=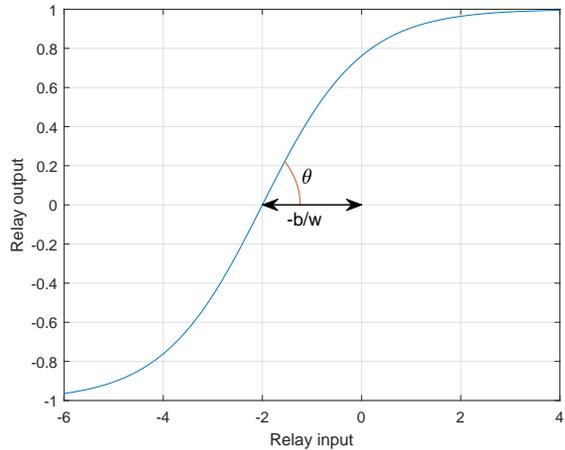,width=0.45\textwidth,trim=0 0 0cm 0, clip}
\caption{\label{fig:tr_func} Transfer function of a relay with a gain of $w$ and a bias of $b$. The slope at the linear regime is determined by the gain such that $w=\tan(\theta)$.}
\end{figure}

 A typical relay transfer function is depicted in Fig. \ref{fig:tr_func}. This transfer function has two controlled parameters: the relay gain, $w$, and an added bias, $b$. (Tradition analysis does not consider the bias as it cannot improve the performance in a linear model.) 

The resemblance of Fig \ref{fig:tr_func} to the common transfer function of a neuron in a neural network leads to the concept of using deep learning tools for the network optimization. Thus, the network is tuned by setting the gain and bias of each relay, and we use back propagation to optimize the network.

 In section \ref{sec:numerical} we consider  specific spatial structures of the network, which determine the channel gains between each two nodes. At this stage, the network is  defined simply by stating the channel gains. We only require that the network has no loops. 
Thus, we can look at the system as a cascade network, in which the relays are divided into layers, and each relay can receive signals only from the input and from relays at previous layers.

Let $y_{i,n} $ be the signal received by relay $n$ in layer $i$, and $\mathbf y_i=[y_{i,1},\ldots,y_{i,N_i}]^T$ be the vector of inputs for all relays in layer $i$ (with $N_i$ being the number of relays in layer $i$).  The received signal vector at layer 1 is given by:
\begin{IEEEeqnarray}{rCl}\label{e:input_first_layer}
\mathbf{y}_1=\mathbf{h}_1 s+\mathbf{n}_1
\end{IEEEeqnarray}
where $ s$ is the signal transmitted by the BS, $\mathbf h_i$ is the vector of channel gains from the BS to the relays of layer $i$, and $\mathbf n_i$ is the vector of additive white noise at each relay, which is assumed to have  independent Gaussian distribution with zero mean and variance $\sigma^2$.

 The signal at relay $n$ of layer $i$ is amplified by a gain of $w_{i,n}$, added to a bias term $b_{i,n}$ and then is subjected to the amplifier non-linearity. Again, without loss of generality, we set the maximal absolute value at the output of each transmitter to be $1$. The non-linearity is represented by a hyperbolic tangent. Thus, the signal at the output of layer $i$ is given by:
\begin{IEEEeqnarray}{rCl}\label{e:relay_output}
\mathbf o_i = \tanh\left(\mbox{diag}(\mathbf w_i)\mathbf y_i+\mathbf b_i\right).
\end{IEEEeqnarray}
The input for layer $i>1$ is given by:
\begin{IEEEeqnarray}{rCl}\label{e:input_general_layer}
\mathbf y_i =\mathbf h_i s +\sum_{\ell=1}^{i-1}\mathbf F_{i,\ell}\mathbf o_\ell+\mathbf n_i 
\end{IEEEeqnarray}
where $\mathbf F_{i,\ell}$ is the matrix of channel gains from layer $\ell$ to layer $i$.
Finally, we assume no direct link from the input to the receivers, and the received signal at user $m$ is:
\begin{IEEEeqnarray}{rCl}\label{e:network_output}
r_m=\sum_{i=1}^d \mathbf g_{i,m}^T\mathbf o_i+\tilde{  n}_m
.
\end{IEEEeqnarray}
where $d$ is the number of layers, $\mathbf g_{i,m}$ is the vector of channel gains from layer $i$ to receiver $m$ and $\tilde{  n}_m$ is the additive Gaussian noise, again with variance $\sigma^2$.

The receiver applies a set of detection functions to produce bit estimates:
\begin{IEEEeqnarray}{rCl}\label{e:network_decisions}
\hat u_{m,k}=q_{m,k}(r_m), \quad m=1,\ldots, M,\ k=1,\ldots,K.\IEEEeqnarraynumspace
\end{IEEEeqnarray}
and the performance is measured by the bit error rate (BER) given by
\begin{IEEEeqnarray}{rCl}\label{e:XXX}
\epsilon_{m,k}=\Pr(\hat u_{m,k}\ne u_{m,k}).
\end{IEEEeqnarray}
We will focus on max-min BER optimization. Thus, our network performance metric will be 
\begin{IEEEeqnarray}{rCl}\label{e:XXX}
\epsilon=\max_{m,k} \epsilon_{m,k}.
\end{IEEEeqnarray}
For low complexity receivers, the choice of detection functions, $q_{m,k}$ should take into account both BER minimization and receiver simplification. For example, for $K=1$ the simplest (and hence preferred) detection function is of the form $\hat u_{m,1}=1$ if $r_m>0$ and $0$ otherwise. 


\section{Traditional (linear model) optimization}\label{Sec:Linear}

The traditional approach treats the relays as linear amplifiers. To that end, we constrain the relay output power to a low enough level, such that the hyperbolic tangent function can be reasonably approximated as linear. 
In mathematical terms, the linear model is obtained by replacing \eqref{e:relay_output} with 
\begin{IEEEeqnarray}{rCl}\label{e:linear_relay}
\mathbf o_i \approx \mbox{diag}(\mathbf w_i)\mathbf y_i+\mathbf b_i
\end{IEEEeqnarray}
and adding a constraint $E[o_{i,n}^2]\le P_{\max}$.

Using \eqref{e:linear_relay} instead of  \eqref{e:relay_output}, the complete network is linear. Thus, each receiver will receive a scaled version of the transmitted signal plus additive Gaussian noise. In such a scenario, BER minimization  is obtained by weighted signal to noise ratio (SNR) maximization subject to the power constraint.  Setting $\mathbf{b}_i=0$ for all $i$,  we need to solve the optimization problem:
\begin{IEEEeqnarray}{rCl}\label{e:linear_opt}
\max_{\{\mathbf{w}_i\}} \min_m &\ &\zeta_m\cdot\mbox{SNR}_m
\notag\\
\text{Subject to:}&& E[o_{i,n}^2]\le P_{\max} \quad i=1... d,\ n=1 ... N_i.
\IEEEeqnarraynumspace
\end{IEEEeqnarray}
where $\mbox{SNR}_m=E\left[\left|[E[r_m|\mathbf{u}_m]\right|^2\right]/\mbox{Var}(r_m|\mathbf{u}_m)$. The SNR weights, $\zeta_m$ are chosen to balance the BER of the different users. 

The optimization problem in  \eqref{e:linear_opt} is not convex, and its solution was not derived so far. The closest solution is the one derived by  Phan et al. \cite{good_phan2012beamforming}. In the journal version of this paper \cite{Bergel_prep}  we extend \cite{good_phan2012beamforming} to solve also the problem at hand. This extension includes alternating minimization over the layers,  rewriting the problem in an efficient manner that reflects the cascade structure and redefining the optimization variables accordingly.

\section{Deep learning optimization}


\subsection{Optimization approach} In our novel approach, we use deep learning training for the optimization of the network. This training allows the relays to use their non-linear regime, as long as the total network performance increases. 

We should note that there is a conceptual difference between the training of the relay network as opposed to neural networks. In the relay case, the network input contains very little information about the desired functionality. For example, in the presented example, we have only $4$ possible network inputs. As the inputs propagate through
the network they accumulate noise, which is the main adversary in this scenario. Thus, by transmitting each of  the few possible inputs many times over the network, we receive many different behaviors and different outputs. 

\subsection{Training}
The network training is based on transmitting pilot symbols over the network and observing the input and output of each relay and the signals received at the receivers. This resembles the training of a neural network that relies on labeled
data to learn its mapping. 

We rewrite the network output, \eqref{e:network_output} as a function of its input and trainable parameters:
\begin{IEEEeqnarray}{rCl}\label{e:XXX}
{\bf r}=\mathfrak{f}({\bf u};{\bm\varphi} )
\end{IEEEeqnarray}
where ${\bf r}=[r_1, \ldots, r_M]^T$ and the trainable parameters are collected into ${\bm \varphi}=[{\bf w}_1^T,\ldots,{\bf w}_d^T, {\bf b}_1^T,\ldots,{\bf b}_d^T]^T$. Note that (unlike most neural networks) the function $\mathfrak f()$ is a random function due to the effect of the noise (see \eqref{e:input_first_layer} and \eqref{e:input_general_layer}). The functionality of $\mathfrak f()$ is determined by the network topology as defined by the channel gains $\mathbf h_{i}$, $\mathbf F_{i,\ell}$ and $\mathbf g_{i,m}$, $i=1,\ldots,d$ $\ell=1,\ldots, d-1$ and $m=1,\ldots, M$.

The network optimization can be performed in a distributed online manner, or, in a centralized batch manner. For brevity, we present here only the centralized version and leave the distributed version to the full version \cite{Bergel_prep}. 

The network training is based on the digital twin paradigm, where the central processor optimizes a simulated version of the network, and then, the optimized parameters  are fed into the actual network. Thus, the centralized optimization requires a central processor that has knowledge of all channel gains, but, does not require actual pilot transmissions except for those needed for the channel estimation.

We collect the data of $B$ (simulated) symbols into a single batch
and use $b$ as the symbol index in the batch. We also apply a normalization stage for each output, that is 
\begin{IEEEeqnarray}{rCl}\label{e:XXX}
\tilde r_m[b] = \frac{r_m[b]}{\sqrt{\frac{1}{B}\sum_{\tilde b=1}^B r_m^2[b]}}
\end{IEEEeqnarray}
and ${\bf \tilde r[b]}=[r_1[b],\ldots,r_M[b]]^T$.
 The training is performed by minimizing the loss  \begin{IEEEeqnarray}{rCl}\label{e:XXX}
\mathcal L({\bm \varphi})=\frac{1}{BMK}\sum_{b=1}^B \sum_{m=1}^M\sum_{k=1}^K L_{m,k}(u_{m,k}[b],\tilde r_m[b])\IEEEeqnarraynumspace
\end{IEEEeqnarray}
where $L_{m,k}(u,\tilde r)$ is the loss function for bit $k$ of user $m$. 

In the numerical results reported below we use $K=1$ and $M=2$ and hence we need two loss functions. User $1$ uses a BPSK receiver. Thus we use 
\begin{IEEEeqnarray}{rCl}\label{e:XXX}
L_{1,1}(u,\tilde r) =\log_2(1+e^{-\tilde r\cdot(-1)^u}).
\end{IEEEeqnarray}
which is obtained from the combination of a sigmoid function to transfer the network output to probability, followed by a log-loss function and a soft-max. The same function is used also for user $2$ when we wish it to use a simple BPSK receiver (and hence, the network is required to perform the data separation for user 2). 

When user $2$ uses a PAM receiver, its data is contained in the absolute value of the input. Thus we use a similar function preceded by a squaring operation with a threshold of $1$ (after normalization). This loss function is given by:
 \begin{IEEEeqnarray}{rCl}\label{e:XXX}
L_{2,1}(u,\tilde r) =\log_2(1+e^{-(\tilde r^2-1)\cdot(-1)^u}).
\end{IEEEeqnarray}

The training module follows a gradient-based minimization of the loss using iterations of the form
\begin{IEEEeqnarray}{rCl}\label{e:XXX}
{\bm \varphi}^{(t+1)}={\bm \varphi}^{(t)}-\eta \nabla_{\bm \varphi}\mathcal{L}({\bm \varphi} ^{(t+1)}).
\end{IEEEeqnarray}
\subsection{Testing and validation}
Note that we do not need to set aside data for testing and validation. These operations are always performed using new random noises (as we can always generate more data).

\subsection{Initialization}
The most direct approach to initialize the training is by using the result of the linear optimization of Section \ref{Sec:Linear}. That is, we use the optimal gains obtained from \eqref{e:linear_opt} and initialize all biases to $\mathbf b_i=\mathbf 0$. But, for large networks, the solution of \eqref{e:linear_opt} requires sequential solution of many convex problems and hence is quite complicated.

Instead, we present here a simpler initialization which showed good performance in our numerical study. The main idea of this initialization is to keep all relays close to their linear regime, so that no relay starts saturated.

We start with layer $i=1$ and then move forward. For relay $n$ of layer $i$ we calculate 
$$p_{i,n}=E[y_{i,n}^2].$$
(Recall that $p_{i,n}$ depends only on the gains and biases of previous layers). Then we draw a random sign $s_{i,n}\in\{-1,1\}$ and random amplitude $ a_{i,n}$ which is uniformly distributed over $[0.5,1] $, and set:
\begin{IEEEeqnarray}{rCl}\label{e:XXX}
w_{i,n}=\frac{a_{i,n}s_{i,n}}{p_{i,n}},\quad b_{i,n}=0.
\end{IEEEeqnarray}
Going over all layers from $i=1$ till $i=d$ we establish a starting point with good dynamic behavior.
 
\begin{figure}[t]
\centering
\begin{minipage}{.48\textwidth}
\centerline{\epsfig{figure=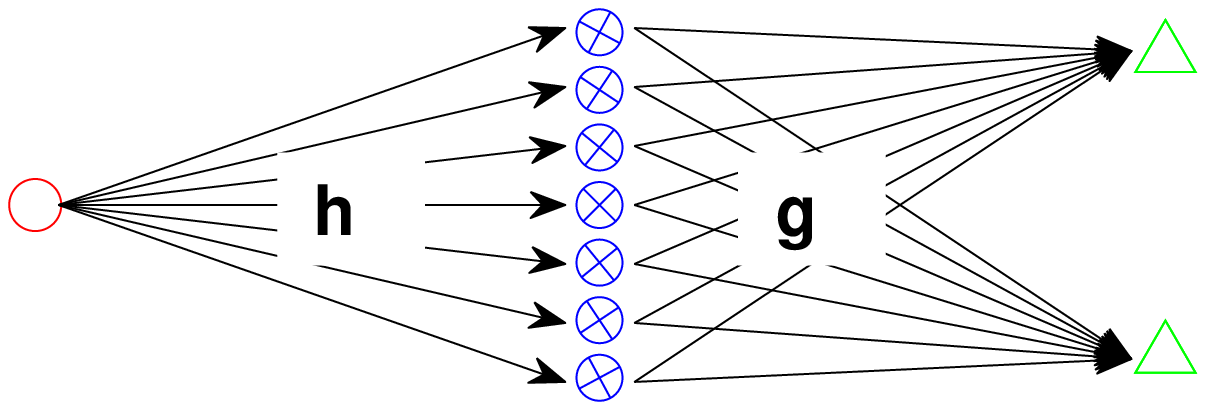,width=.95\textwidth}}
\caption{\label{fig:1layer} A relay network with 7 relays in a single layer.}
\end{minipage}%
\begin{minipage}{.1\textwidth}
\phantom{aa}
\end{minipage}
\begin{minipage}{.48\textwidth}
  \centering
\centerline{\epsfig{figure=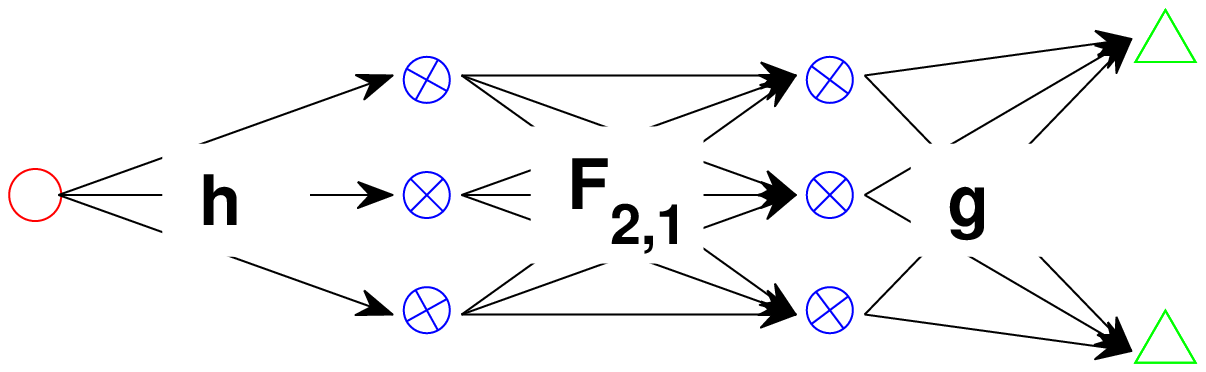,width=.95\textwidth,trim=0 0 0cm 0cm, clip}}
\caption{\label{fig:2layers} A relay network with 2 layers each with 3 relays.}
\end{minipage}
\end{figure}

\section{Numerical results}\label{sec:numerical}

\subsection{Simple networks}
In this section we demonstrate the advantages of our novel approach. We start by studying simple networks with one and two layers. 

\begin{figure}[t]
\centerline{\epsfig{figure=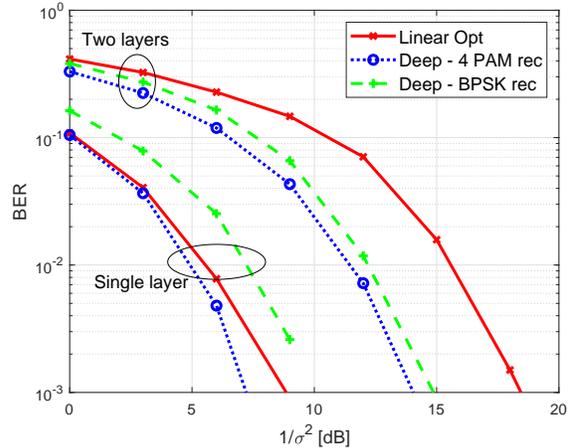,width=.45\textwidth}}
\caption{\label{fig:BER_1_2_layers} BER vs. $1/sigma^2$ for the relay networks of Fig. \ref{fig:1layer} and Fig. \ref{fig:2layers}. The figure compares traditional (linear) optimization and deep learning optimization.}
\vspace{-3mm}
\end{figure}

Starting with the single layer network of Fig. \ref{fig:1layer}, we use $\mathbf{h}_1=[3,1,-1,-3,-1,1,3]^T$ and $\mathbf g_{1,1}=\mathbf g_{1,2}=[1,1,1,1,1,1,1]^T$. The resulting BER is depicted in Fig. \ref{fig:BER_1_2_layers} as a function of $1/\sigma^2$.
The figure shows the BER of the worst user in $3$ scenarios. The solid line depicts the BER in the traditional approach, where the optimization treats the relays as linear amplifiers with a power constraint (we set $P_{\max}=0.64$). The dotted line shows the result of a deep learning training. Here, we keep the original receiver structure, that is, user $1$ employs a BPSK receiver, while user $2$  employs a 4 PAM receiver according to Table I. The shallow nature of this  network allows the deep learning optimization to gain only $0.7$dB over the traditional optimization. This gain is obtained by a slight increase of the relays gain at the price of some non-linear response (which is basically equivalent to an adaptive tuning of $P_{\max}$ in \eqref{e:linear_opt}).     

The deep learning approach allows us also to simplify the receivers, and use BPSK receivers for both users. In such a case, the network is responsible for the data separation so that user $2$ will not be affected by the data of user $1$. In this shallow network, we see that this data separation is feasible, but it comes with a cost of more than $1.5$dB.

We next consider the (slightly) more complicated 2 layer network of Fig. \ref{fig:2layers}. Here we use: ${\bf h}_1=[2,2,2]^T$,  ${\bf h}_2={\bf 0}$, $\mathbf g_{2,1}=[0,4,2]^T$, $\mathbf g_{2,2}=[2,4,0]^T$, $\mathbf g_{1,1}=\mathbf g_{1,2}=\mathbf{0}$ and 
\begin{IEEEeqnarray}{rCl}\label{e:XXX}
\mathbf{F}_{2,1}=\begin{bmatrix}1 &-.5& 1\\-.5&-1&-.5\\1 &-.5& 1\end{bmatrix}.
\end{IEEEeqnarray}

The resulting BER is also depicted in Fig. \ref{fig:BER_1_2_layers}. In this network our novel approach bring a gain of $4$dB over the traditional approach. This gain is significant as we can reduce all transmission powers by a factor of more than $2$ at the only price of smarter optimization. Furthermore, the data separation (using only BPSK receivers) comes here at a negligible cost of less than $0.5$dB.

\subsection{Spatial distribution of relays}
We next consider a more practical model, where the relays are distributed over a plane. We consider a network with one single antenna
base station (BS) transmitter, two single antenna receivers and 100 relays with directional
antennas, as illustrated in Fig. \ref{fig:system_model}. We consider a sector of $60^\circ$ in a cell with a radius of 20.
The receivers are located at the cell edge. To prevent loops in the network, we assume that each relay is equipped with two directional antennas, each of $90^\circ$ width. The receive antenna is pointing backward to the transmitter (and will receive also any
relay in its beam width). The transmit antenna is pointing to the opposite direction (forward). 
(In Fig. \ref{fig:system_model}, the sectors within each relay represent the beams of the directional antennas.)

We assume Gaussian fading over
all channels, so that each channel gain is given by $r^{-\alpha}\cdot u$ where r is the link length, $\alpha = 4$ is
the path loss exponent and $u$ is an independent Gaussian fading variable with zero mean and unit variance.

\begin{figure}[t]
\centerline{\epsfig{figure=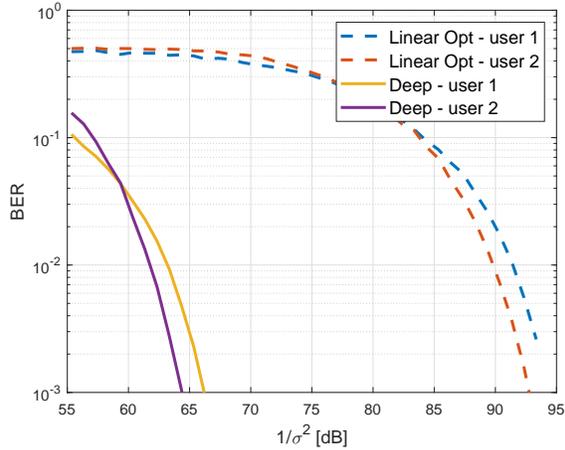,width=.45\textwidth}}
\caption{\label{fig:BER} BER vs. $1/\sigma^2$ for a cellular network with 100 relays. The figure compares traditional (linear) network optimization vs. deep relay optimization.}
\end{figure}

The  resulting BER curves after optimization for both users are depicted in Fig. \ref{fig:BER}.
Here the better optimization approach produced huge gains which exceeds $25$dB. Furthermore, the cost of data separation is completely negligible (and hence we present only the BER with the simpler (BPSK) receivers for both users). 

\begin{figure}[t]
\centerline{\epsfig{figure=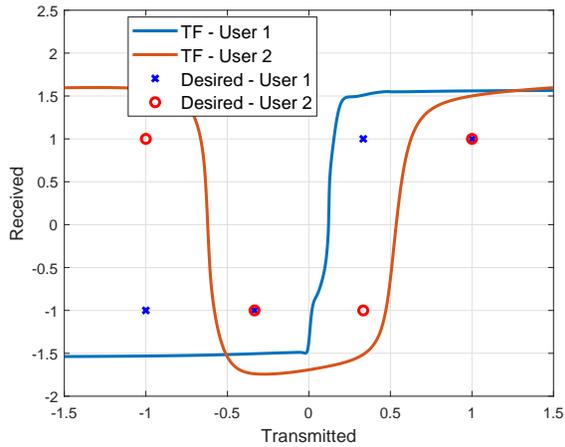,width=.45\textwidth}}
 \caption{\label{fig:BER_transfer}  Transfer functions of the relay network for the two users. x-marks and circles show the desired values used for network training.}
\end{figure}

Obviously, the presented BER  curves prove that network managed to learn the proper transfer function to allow the data separation. To better demonstrate this, Fig. \ref{fig:BER_transfer} depicts the  transfer functions (input to output) of the relay network to both users (without noise). One can see that the transfer function for the second user is indeed separating the desired bits while ignoring the data of user 1.

\section{Conclusions}
This paper presented a novel approach for the optimization of relay networks. Unlike the traditional approach that approximates relay as linear amplifiers, our novel approach takes into account the true non-linear nature of the relays. Using the similarity between the transfer function of a relay and the transfer function of a neuron, we employ deep learning methodology to better optimize the network. Numerical study shows huge gains  compared to traditional optimization.

\bibliographystyle{ieeetr}
\bibliography{../DR_downlink_paper/DR_downlink}

\begin{thebibliography}{10}

\bibitem{harmon1912girdling}
D.~Harmon, ``Girdling the globe by wireless,'' {\em Scientific American},
  vol.~106, no.~16, pp.~360--370, 1912.

\bibitem{cover1979capacity}
T.~Cover and A.~E. Gamal, ``Capacity theorems for the relay channel,'' {\em
  IEEE Transactions on information theory}, vol.~25, no.~5, pp.~572--584, 1979.

\bibitem{laneman2004cooperative}
J.~N. Laneman, D.~N. Tse, and G.~W. Wornell, ``Cooperative diversity in
  wireless networks: Efficient protocols and outage behavior,'' {\em IEEE
  Transactions on Information theory}, vol.~50, no.~12, pp.~3062--3080, 2004.

\bibitem{nabar2004fading}
R.~U. Nabar, H.~Bolcskei, and F.~W. Kneubuhler, ``Fading relay channels:
  Performance limits and space-time signal design,'' {\em IEEE Journal on
  Selected Areas in communications}, vol.~22, no.~6, pp.~1099--1109, 2004.

\bibitem{zhao2006improving}
Y.~Zhao, R.~Adve, and T.~J. Lim, ``Improving amplify-and-forward relay
  networks: optimal power allocation versus selection,'' in {\em 2006 IEEE
  international symposium on information theory}, pp.~1234--1238, 2006.

\bibitem{nasir2013relaying}
A.~A. Nasir, X.~Zhou, S.~Durrani, and R.~A. Kennedy, ``Relaying protocols for
  wireless energy harvesting and information processing,'' {\em IEEE
  Transactions on Wireless Communications}, vol.~12, no.~7, pp.~3622--3636,
  2013.

\bibitem{lu2014wireless}
X.~Lu, P.~Wang, D.~Niyato, D.~I. Kim, and Z.~Han, ``{Wireless networks with RF
  energy harvesting: A contemporary survey},'' {\em IEEE Communications Surveys
  \& Tutorials}, vol.~17, no.~2, pp.~757--789, 2014.

\bibitem{riihonen2011mitigation}
T.~Riihonen, S.~Werner, and R.~Wichman, ``Mitigation of loopback
  self-interference in full-duplex {MIMO} relays,'' {\em IEEE transactions on
  signal processing}, vol.~59, no.~12, pp.~5983--5993, 2011.

\bibitem{sabharwal2014band}
A.~Sabharwal, P.~Schniter, D.~Guo, D.~W. Bliss, S.~Rangarajan, and R.~Wichman,
  ``In-band full-duplex wireless: Challenges and opportunities,'' {\em IEEE
  Journal on selected areas in communications}, vol.~32, no.~9, pp.~1637--1652,
  2014.

\bibitem{good_phan2012beamforming}
A.~H. Phan, H.~D. Tuan, H.~H. Kha, and H.~H. Nguyen, ``Beamforming optimization
  in multi-user amplify-and-forward wireless relay networks,'' {\em IEEE
  transactions on wireless communications}, vol.~11, no.~4, pp.~1510--1520,
  2012.

\bibitem{sanguinetti2012tutorial}
L.~Sanguinetti, A.~A. D'Amico, and Y.~Rong, ``A tutorial on the optimization of
  amplify-and-forward {MIMO} relay systems,'' {\em IEEE Journal on Selected
  Areas in Communications}, vol.~30, no.~8, pp.~1331--1346, 2012.

\bibitem{chen2019artificial}
M.~Chen, U.~Challita, W.~Saad, C.~Yin, and M.~Debbah, ``Artificial neural
  networks-based machine learning for wireless networks: A tutorial,'' {\em
  IEEE Communications Surveys \& Tutorials}, vol.~21, no.~4, pp.~3039--3071,
  2019.

\bibitem{nikbakht2020unsupervised}
R.~Nikbakht, A.~Jonsson, and A.~Lozano, ``Unsupervised learning for parametric
  optimization,'' {\em IEEE Communications Letters}, vol.~25, no.~3,
  pp.~678--681, 2020.

\bibitem{sholev2020neural}
O.~Sholev, H.~H. Permuter, E.~Ben-Dror, and W.~Liang, ``Neural network {MIMO}
  detection for coded wireless communication with impairments,'' in {\em 2020
  IEEE Wireless Communications and Networking Conference (WCNC)}, pp.~1--8,
  2020.

\bibitem{saxena2021reinforcement}
V.~Saxena, H.~Tullberg, and J.~Jald{\'e}n, ``Reinforcement learning for
  efficient and tuning-free link adaptation,'' {\em IEEE Transactions on
  Wireless Communications}, vol.~21, no.~2, pp.~768--780, 2021.

\bibitem{zhang2020neural}
Z.~Zhang, Y.~Lu, Y.~Huang, and P.~Zhang, ``Neural network-based relay selection
  in two-way {SWIPT}-enabled cognitive radio networks,'' {\em IEEE Transactions
  on Vehicular Technology}, vol.~69, no.~6, pp.~6264--6274, 2020.

\bibitem{xu2021intelligent}
L.~Xu, X.~Yu, and T.~A. Gulliver, ``Intelligent outage probability prediction
  for mobile iot networks based on an igwo-elman neural network,'' {\em IEEE
  Transactions on Vehicular Technology}, vol.~70, no.~2, pp.~1365--1375, 2021.

\bibitem{guo2021energy}
Y.-Y. Guo, J.~Yang, X.-L. Tan, and Q.~Liu, ``An energy-efficiency multi-relay
  selection and power allocation based on deep neural network for
  amplify-and-forward cooperative transmission,'' {\em IEEE Wireless
  Communications Letters}, vol.~11, no.~1, pp.~63--66, 2021.

\bibitem{Bergel_prep}
I.~Bergel, ``Non-linear relay optimization using deep learning tools,'' {\em in
  preparation}, 2023.

\end{thebibliography}

\end{document}